\documentclass[preprint,prb, superscriptaddress, showkeys, showpacs]{revtex4-1}

\usepackage{amsmath}    
\usepackage{graphicx}   
\usepackage{subfigure}  
\usepackage{hyperref}   
\usepackage{bm}		
\usepackage{float}

\preprint{\bf PREPRINT}

\begin{document}
\draft

\title{Using low moments of the Liouvillian to calculate mode
  lifetimes in low dimensional models}
\author {Yang Gao}
\affiliation{Department of Physics $\&$ Astronomy, Clemson University,
  Clemson, SC 29634}
\author {Doyl Dickel}
\affiliation{Department of Physics $\&$ Astronomy, Clemson University,
  Clemson, SC 29634}
\affiliation{Now at Karlsruhe Institute of Technology, Institute of Applied 
Materials (IAM-ZBS), Kaiserstr. 12, 76131 Karlsruhe}
\author {David Harrison}
\affiliation{Department of Physics $\&$ Astronomy, Clemson University,
  Clemson, SC 29634}
\affiliation{Now at Department of Physics, Wake Forest University,
  Winston-Salem, NC}
\author {Murray S. Daw}
\affiliation{Department of Physics $\&$ Astronomy, Clemson University,
  Clemson, SC 29634}



\begin{abstract}
  A recent proposal~\cite{DD1,DD2} for practical calculation of
  vibrational mode lifetimes is tested on simple, low-dimensional
  anharmonic models. The proposed scheme approximates the mode
  lifetime in terms of ensemble averages of specific functions in
  phase-space; various levels of approximation correspond to ensemble
  moments of the Liouvillian. It is shown that, for systems where the
  vibrational density of states is well-approximated by a single
  broadened peak, the fourth-moment approximation works well over the
  full range of temperature.
\end{abstract}

\maketitle

\section{Introduction}

Dickel \& Daw~\cite{DD1, DD2} recently proposed an efficient,
approximate means of calculating vibrational mode lifetimes in
solids. The method involves ensemble averages of appropriate functions
in phase space that can be carried out by conventional Monte Carlo in
combination with a means of calculating forces, such as interatomic
potentials or first-principles electronic structure codes. The
approach was illustrated on a lattice model of non-linear
interactions, where the dependence of the mode lifetimes on cell size
and temperature was investigated numerically. 

While the aim of the original work was to further calculations of
vibrational mode lifetimes in solids, the purpose of the present work
is to examine in more detail the approximations involved in the
method. To this end we take up the same method as applied to very
simple systems of just one or two degrees of freedom. In considering
systems of such simplicity, we analyzed some aspects of the problem
analytically as well as numerically, and the insights obtained are
reported here. These insights are expected to prove fruitful in the
application of this method to the original target (vibrational
lifetimes in solids).

This paper is organized as follows. First, we recap briefly the
proposal of Dickel \& Daw (DD). Then we apply the
proposed method to the simple dynamical models considered here. Our
analysis of the results focuses on the density of states, by which we
can understand when and why the approximations work as they
do. Finally, we draw our conclusions.

\section{Background \& Scope of the Present Work}

We summarize here the proposed method of DD, who began by examining
the momentum Auto-Correlation Function (MACF)
\begin{equation}
\chi_p (t) = \frac{\langle  p(0) p(t) \rangle}{\langle
   p^2 \rangle}  
\label{eq:chioft}
\end{equation}
where the angular brackets indicate phase-space averages over the
canonical ensemble at temperature $T$ ($\rho = \exp{(-H/T)}$). 

The auto-correlation can be studied in terms of the
Liouvillian~\cite{Koopman31,Koopman32}, which 
governs the time evolution of functions $f(x,p,t)$ in phase space
according to
\[ \frac{\partial f}{\partial t} = -i \hat{L} f \]
where the (hermitian) Liouvillian operator is
\[ \hat{L} = i\{H, \} = 
i \sum_l (
\frac{\partial H}{\partial x_l} \frac{\partial}{\partial p_l} - 
\frac{\partial H}{\partial p_l} \frac{\partial}{\partial x_l} )
\]
The equation of motion can be integrated formally, so that 
\[ f(x,p,t) = e^{-i t \hat{L}} f(x,p,0) \] 
and we can express the auto-correlation explicitly in
terms of $\hat{L}$:
\[
\chi (t) = \frac{\langle p e^{-i t \hat{L}} p \rangle}{\langle p^2 \rangle} 
\]

The Taylor Series of $\chi(t)$ 
\[
\chi(t) = 1- \mu_2 \frac{t^2}{2!} + \mu_4 \frac{t^4}{4!}- \mu_6 \frac{t^6}{6!}+... 
\]
relates the derivatives of $\chi(t)$ at $t=0$ to the moments of the
Liouvillian acting on the momentum: 
\[
\mu_n=\frac{\langle p \hat{L}^n p\rangle}{\langle p^2\rangle}
\]
These moments are also the moments of the density of states (DOS)
derived from $\chi(t)$. That is, taking the Fourier transform of
$\chi(t)$ to get $n(\omega)$, the moments are also 
\[
\mu_m=\int_{-\infty}^{+\infty} d\omega\ \omega^m\ n(\omega) 
\]

Auto-correlation functions considered in this work typically have
strong oscillations dampened by some sort of decaying envelope (for
examples, see Figs.~\ref{figmacfx4}-\ref{figmacf_cubic}). We
propose here to use the area under the square of the MACF as a measure
of the lifetime 
\begin{equation}
\tau = \int_{-\infty}^{+\infty} dt\ \chi(t)^2
\label{eq:taudef}
\end{equation}
This is not intended to correspond to a particular physical
measurement that might be performed, but rather is suggested as a
simple generic measure of the rate of the decay of the
correlation. Such a measure also lends itself easily to analysis, as
we shall see. Using Parseval's Theorem, the lifetime is also given as
the area under the $n(\omega)^2$ curve:
\begin{equation}
\tau = \int_{-\infty}^{\infty} dt\ \chi(t)^2 = \int_{-\infty}^{\infty}
d\omega\ n(\omega)^2 
\label{eq:n2}
\end{equation}
DD observed that the lifetime $\tau$ can be expressed as a
function of the moments
\[ \tau = F( \mu_2, \mu_4, \mu_6, \ldots ) \]
which can be re-expressed (using dimensional analysis) as
\[
\tau/\tau_2 = G( \gamma_4, \gamma_6, \ldots)
\]
where $\tau_2 = \mu_2^{-1/2} $ and the $\gamma$'s are dimensionless
parameters 
\[
\gamma_n = \frac{\mu_n}{(\mu_2)^{n/2}} 
\]
that characterize the shape of the DOS. While it is not generally
possible to know all of the moments, DD proposed that in certain
circumstances the lifetime might be practically approximated from a
knowledge of only the lowest moments. This suggests a series of
approximations, starting with only the second moment
\begin{equation} 
\tau = c \tau_2
\label{eqsecond}
\end{equation}
and including successively
higher moments. The fourth moment approximation would then be 
\begin{equation}
\tau = \tau_2 F(\gamma_4)
\label{eqfourth}
\end{equation}
where $F$ is some function yet to be determined. The higher moments
correspond to ensemble averages of higher powers of the Liouvillian,
and so each higher moment involves higher time derivatives of the
dynamical variables. 

DD then went on (in part 2) to test the lowest approximation on a
simple model of non-linear lattice vibrations as a function of cell
size and temperature. First, much as done by Ladd, \emph{et al.},~\cite{Ladd86} DD
calculated from ordinary molecular dynamics the auto-correlation
function for each normal mode a periodic cell of a given size
(appropriately sampled at the specified temperature) and from there
the lifetime. Second, they calculated using standard Monte Carlo the
second moment $\mu_2$ (hence $\tau_2$) for each mode. (This second
part of the demonstration is, of course, requires much less
computational time than the first.) They then plotted $\tau/\tau_2$
vs. temperature for all modes, and found that at high temperatures the
lifetime was simply proportional to $\tau_2$. Furthermore, at high
temperature, the auto-correlation functions scaled in a simple
way. That is, plotting all of the calculated $\chi$ vs. $t/\tau_2$
exhibited a data collapse, revealing that indeed the high-temperature
dynamics of the mode decay could simply be described by a single
parameter. Thus, the high temperature behavior was well approximated at
the lowest level (second moment).

DD ended by speculating that the behavior over the full range of
temperature might be accounted for by including fourth moment, but
that was not tested. Also, that paper did not offer much insight as to
why the second moment approximation should work well at high
temperature but be insufficient at low temperatures.

The present study uses several simple dynamical models as the basis
for testing the fourth moment approximation and also in using the
density of states to provide an analysis of why the approximation
might work and when it would be expected to fail.

\section{Models Considered}
We consider three simple model hamiltonians in one ($x$) and two ($x$
and $y$) dimensions. These models are chosen because they are simple,
non-linear, and the ensemble averages can be obtained
analytically. The momentum conjugate to $x$
is $p$; that to $y$ is $q$. \\

\noindent \underline{\emph{$x^4$ model:}}
\begin{equation} H(p,x) = p^2 + x^2 + x^4 \end{equation}

The auto-correlation in the $x^4$ model has been studied extensively
before~\cite{SenPRL,SenPRE}. In that work, an analytic approximation
to $\chi(t)$ was obtained at low temperature:
\begin{equation}
\chi(t)=\frac{\cos(t)-3 T t \sin(t)}{ 9 T^2 t^2+ 1}
\label{Sen_eq}
\end{equation} 
showing an oscillatory behavior with an \emph{algebraically} decaying
envelope.  Our calculated auto-correlation conforms well to this
analytical form at low temperatures. \\

\noindent \underline{\emph{ $x^2 y^2$ model:}}
\begin{equation} H(p,x,q,y) = p^2 + q^2/M + x^2 + y^2 + x^2 y^2 \end{equation}

The $x^2 y^2$ model is a simple extension to two modes coupled
nonlinearly. In this model, we investigate various values of the ratio
($M$) of the masses between the two modes, which controls the degree
of resonance. \\

\noindent \underline{\emph{``cubic'' model:}}
\begin{equation} H(p,x,q,y) = p^2 + q^2 + x^2 + y^2 + \frac{\lambda}{4}(x^2+y^2)^2
  +\frac{1}{3} (x^3-3xy^2) \end{equation}

The ``cubic'' model for certain parameters has multiple minima in the
$x-y$ plane and exhibits a ``structural'' transformation (from
multiple attractors to a single attractor) with temperature, which
makes it interesting to include in the present study. To explore the
effects produced by this transition, we tried various values of the
strength ($\lambda$) of the cubic term. For $\lambda < 2/9$, there are
3 off-center global minima with one local minimum on-center. For $\lambda
> 1/4$ only there is only 1 global minimum on-center. 

\begin{figure}[H]
\centering
\includegraphics[scale=1.1]{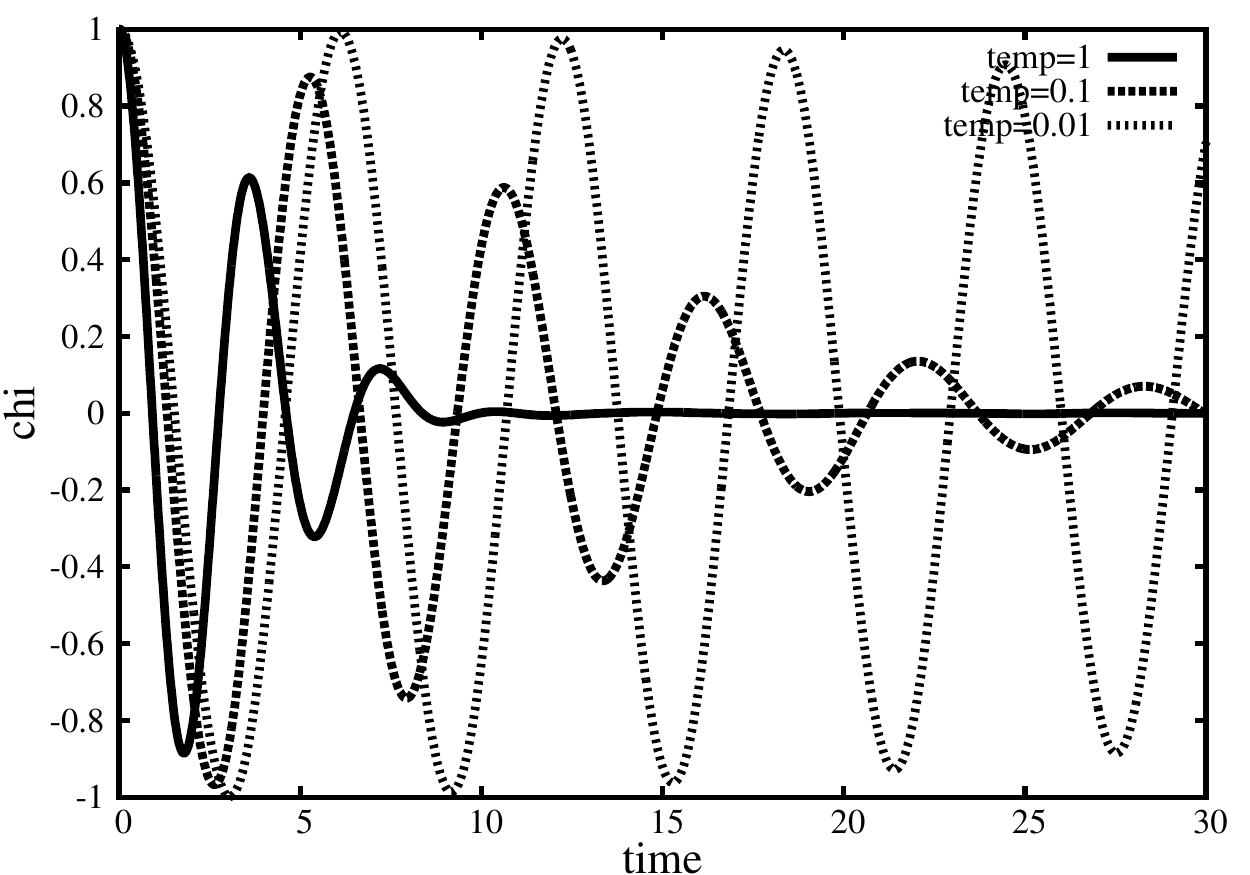}
\caption{The MACF at three temperatures for the $x^4$ model.}
\label{figmacfx4}
\end{figure}

\begin{figure}[H]
\centering
\includegraphics[scale=1.1]{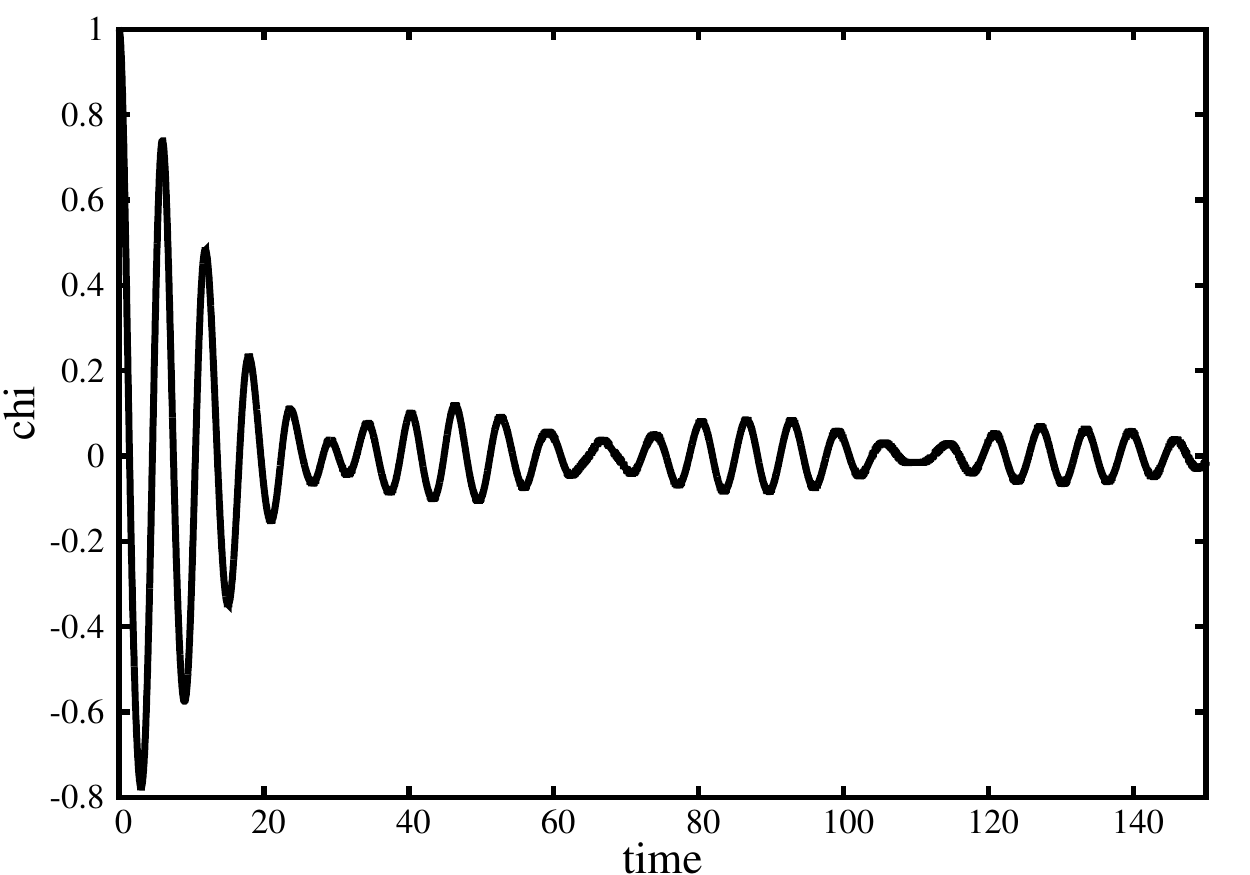}
\caption{The MACF of the $x$-mode at $\lambda=0.5$ and $T=0.2$ for the
  ``cubic'' model.}
\label{figmacf_cubic}	
\end{figure}

Some examples of a calculated MACF are shown in
Figs.~\ref{figmacfx4}-\ref{figmacf_cubic}. For the $x^4$ model, the
function exhibits a simple oscillation and decay. In the ``cubic
model'', the function displays less regularity because of the less
symmetrical potential.

\section{Testing the Fourth Moment Approximation}

\begin{figure}[H]
\centering
\includegraphics[scale=1.1]{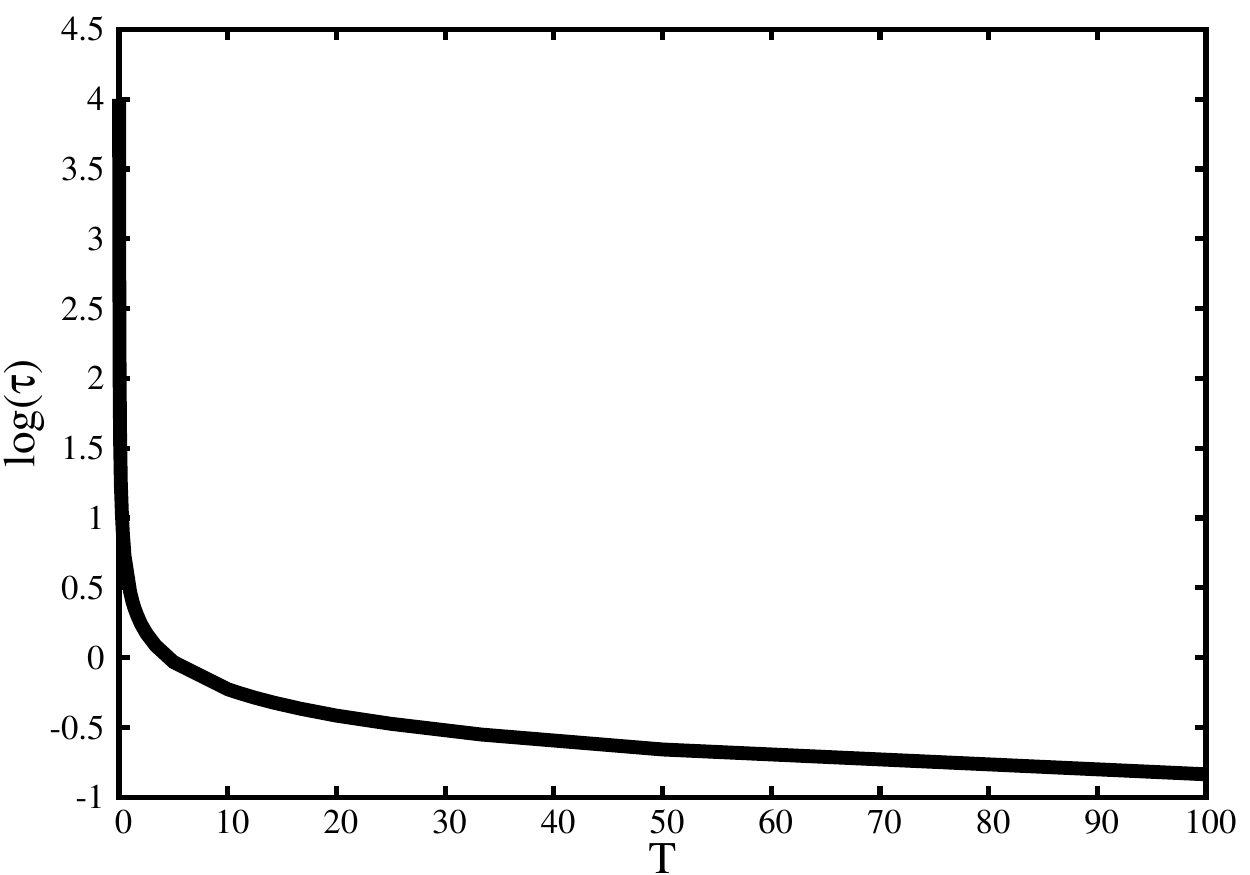}
\caption{Lifetime (Eq.~\ref{eq:taudef}) vs. temperature in the $x^4$ model. }
\label{lifetime_x4}	
\end{figure}

We want to determine if the form in Eq.~\ref{eqfourth} is robust
enough to approximate the lifetimes in the various simple models we
have chosen. In the $x^4$ model, for example, we can perform ensemble
dynamics at various temperatures and extract the lifetime by
Eq.~\ref{eqfourth}. The lifetime vs. temperature for the $x^4$ model
is then shown in Fig.~\ref{lifetime_x4}.

\begin{figure}[H]
\centering
\includegraphics[scale=0.3]{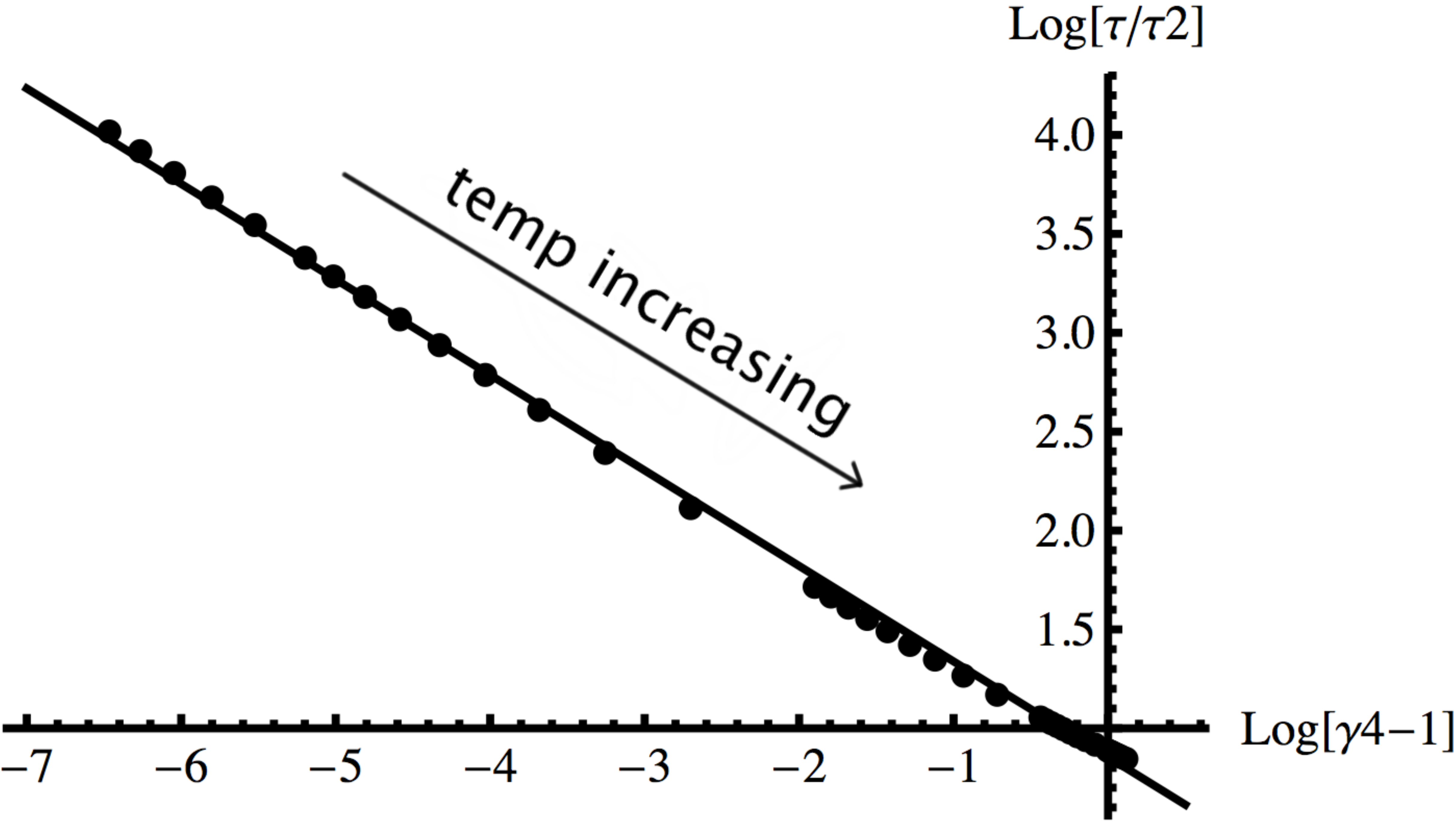}
\caption{Scatterplot of $\tau/\tau_2$ vs $\gamma_4$ for the $x^4$
  model. The straight line is a fit using Eq.~\ref{eqfit}.}
\label{MACFx4}	
\end{figure}

In view of Eq.~\ref{eqfourth}, we represent these results as a
scatterplot of $\tau/\tau_2$ vs. $\gamma_4$, where the
temperature-dependent $\tau_2$ and $\gamma_4$ are calculated
analytically. Noting that $\gamma_4 \geq 1$, and the power-law
behavior of $\tau$ and the moments with $T$, we will plot
$\log(\tau/\tau_2)$ vs. $\log(\gamma_4-1)$. This is shown in
Fig.~\ref{MACFx4}

\begin{figure}[H]
\centering
\includegraphics[scale=1.1]{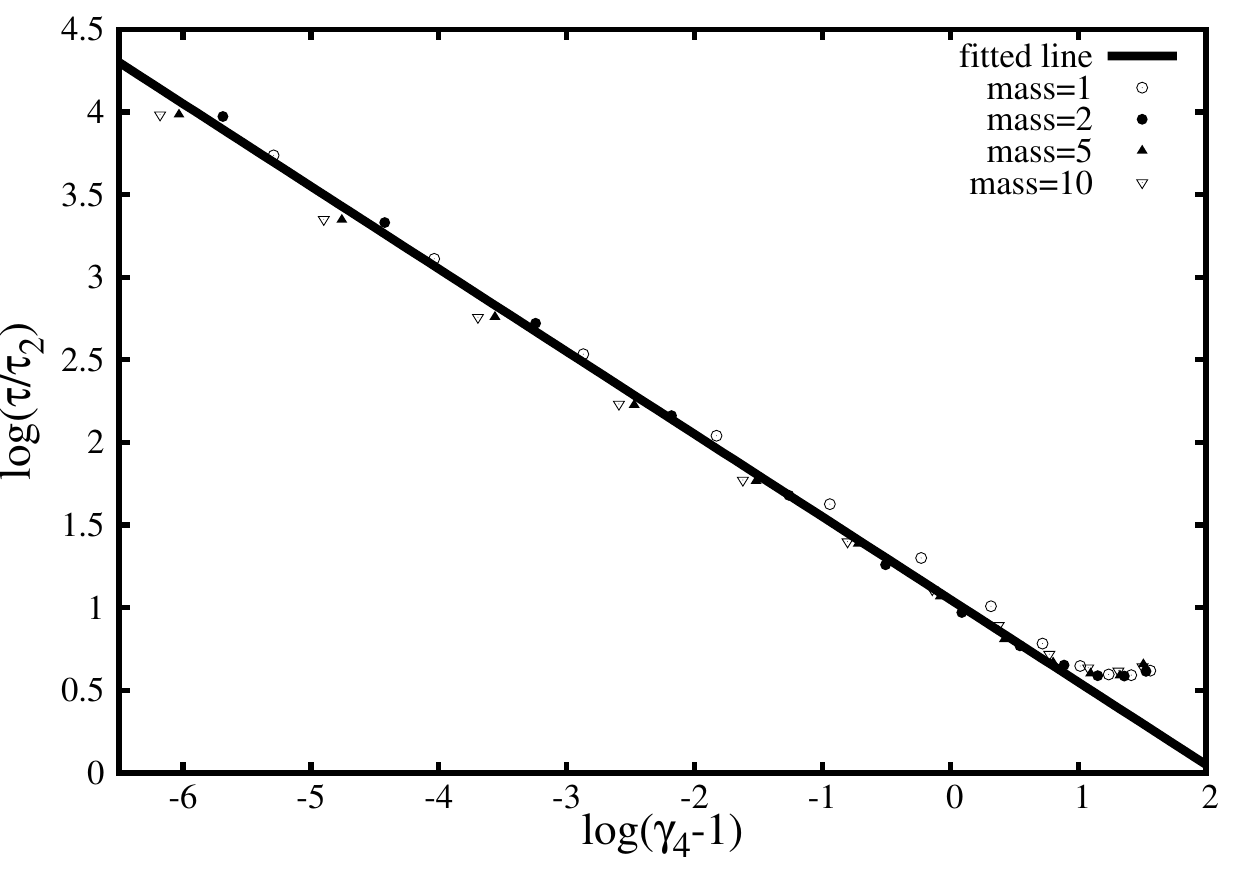}
\caption{$\tau/\tau_2$ vs $\gamma_4$ of the $x$-mode for the $x^2y^2$
  model.  The straight line is a fit using Eq.~\ref{eqfit}. }
\label{MACFx2y2}	
\end{figure}

Similar results can be seen for the $x^2 y^2$ model (see
Fig.~\ref{MACFx2y2}). 

The relations in Figs.~\ref{MACFx4}-\ref{MACFx2y2} are fit well by
\begin{equation}
\tau/\tau_2 \propto ( \gamma _4-1 )^{-1/2}
\label{eqfit}
\end{equation} 
From the analytic approximation to MACF for the $x^4$ model at low
temperature (Eq.~\ref{Sen_eq}), we can also calculate $\tau$, $\mu_2$, and
$\mu_4$, and we find
\begin{equation}
\tau/\tau_2 = \pi (\gamma _4-1)^{-1/2}
\end{equation} 
The analytical form for $\chi(t)$ was derived by Sen, \emph{et al.}
only for low temperatures, by noting the dependence of the oscillator
frequency on energy and the contributions of different energies in
the canonical ensemble. However, here we find the relationship between
$\tau$, $\tau_2$, and $\gamma_4$ extends over a large range of
temperature. The reason for this extended range will be understood
better below.

At low $T$, the low moments for both $x^4$ and $x^2 y^2$ models behave
similarly, in that $\mu_2 \approx 1 + a T$ and $\mu_4 \approx 1 + 2 a T$
so that $\gamma_4$ approaches 1 as $T^2$. Thus $\tau_2$ is approaches
a constant while $\gamma_4-1$ goes to zero, and the lifetime diverges
like $\tau \approx T^{-1}$ at low temperature. The temperature
dependence at low $T$ is dominated by the approach of $\gamma_4$ to
1. 

At high temperature, the moments for the $x^4$ model go as $\mu_2
\approx a T^{1/2} + b$ and $\mu_4 \approx c T - d T^{1/2}$. So
$\gamma_4$ saturates to a constant as $T^{-1/2}$, leaving only the
variation in $\tau_2$ to account for the change in lifetime. Thus the
lifetime at high $T$, is governed by the behavior of $\tau_2$ and
$\tau \approx T^{-1/4}$.

This accounts well for the two power-law regimes visible in
Fig.~\ref{lifetime_x4}. 

For the $x^2 y^2$ model, by contrast, at high temperature, the moments go
as $\mu_2 \approx 2 T/\log{T} + 1/2$ and $\mu_4 \approx 4 T^2/\log{T}
+ 4 T/M$ which makes $\gamma_4$ go as $\log{T}$. This cancels a
$\log{T}$ dependence in $\tau_2$ leaving $\tau \approx T^{-1/2}$. 

\begin{figure}[H]
\centering
\includegraphics[scale=1.1]{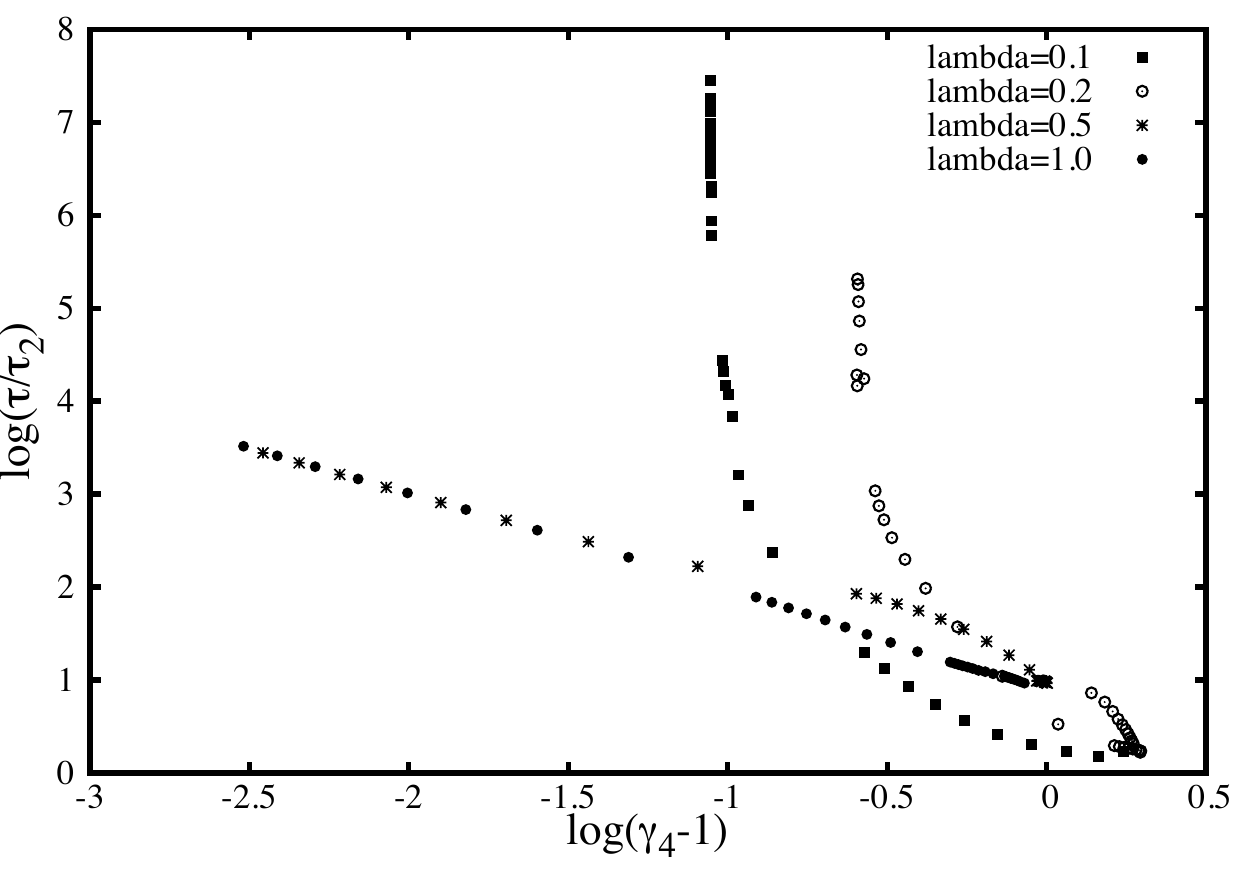}
\caption{Scatterplot of $\tau/\tau_2$ vs $\gamma_4$ of the $x$-mode
  for the ``cubic" model, showing irregular behavior as compared to
  the other models (Figs.~\ref{MACFx4}-\ref{MACFx2y2}).} 
\label{MACFcubic}	
\end{figure}

The $x^4$ and $x^2 y^2$ models seem to be well-described by the simple
combination of the first two moments. However, by contrast, the
corresponding scatterplot for the ``cubic model'' deviates
significantly (Fig.~\ref{MACFcubic}), so that there is no simple
relationship between $\tau$ and the first two moments. Evidently,
higher moments will be required to capture the dynamical behavior of
the cubic model over a wide range of temperatures and parameters.

\section{Analysis}

From the previous results, we see that the behavior of the lifetime
for the $x^4$ and $x^2 y^2$ models over a wide range of parameters
and temperature is captured in the behavior of the two lowest moments
($\mu_2$ and $\mu_4$) which can be calculated analytically. However,
for the cubic model, the behavior is more complex, requiring at least
higher moments in the description. We investigate here the reasons for
success in one case and not in the other.

\begin{figure}[H]
\centering
\includegraphics[scale=1.1]{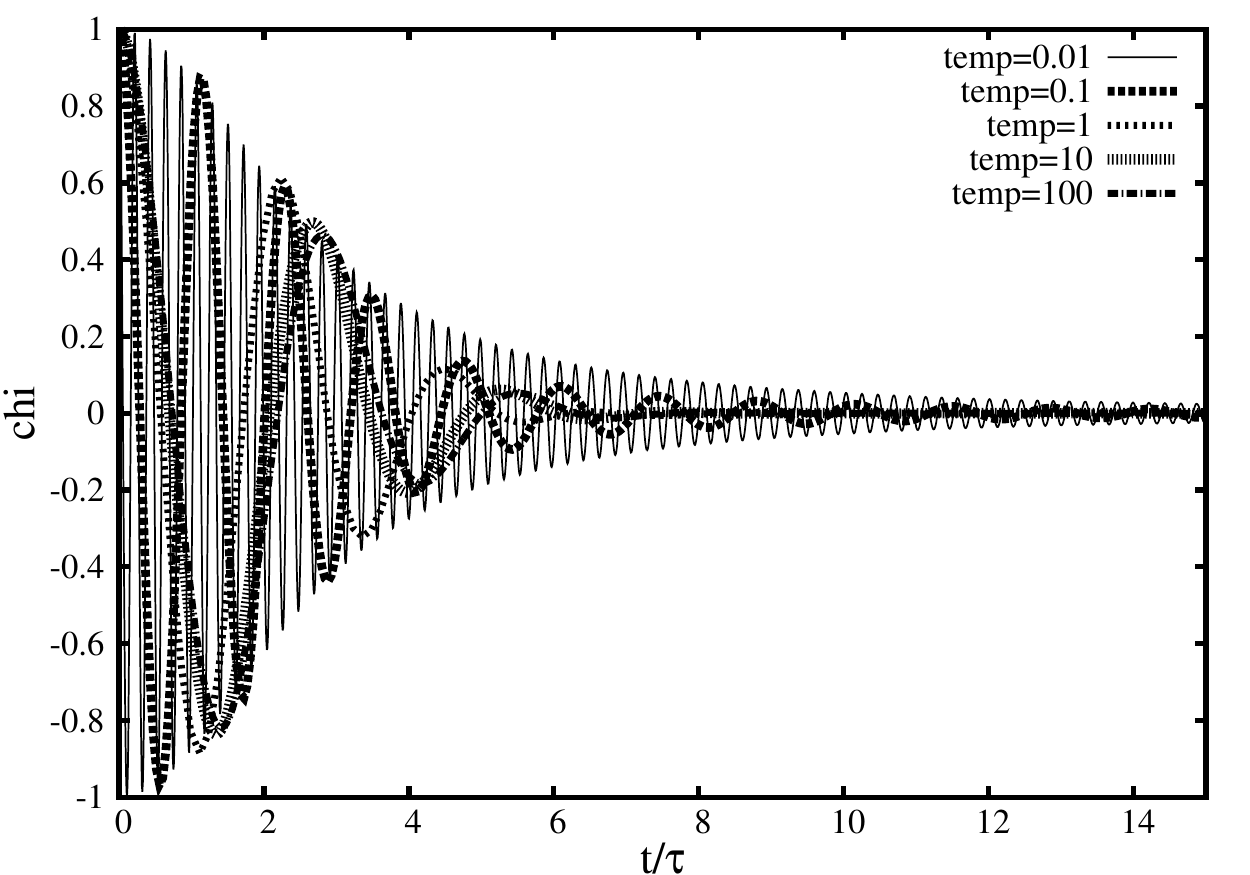}
\caption{Data collapse of MACF for the $x^4$ model (as explained in the text.) }
\label{data_collapse}	
\end{figure}

Fig.~\ref{data_collapse} shows the insight gained from checking for a
data-collapse for $\chi(t)$, by scaling the time $t$ by the lifetime
$\tau$ (Fig.~\ref{lifetime_x4}) for the $x^4$ model. The results
illustrate that while the oscillations of auto-correlation functions
vary with temperature, they are contained by one decaying envelope,
which is what we are trying to capture.

\begin{figure}[H]
\centering
\includegraphics[scale=1.1]{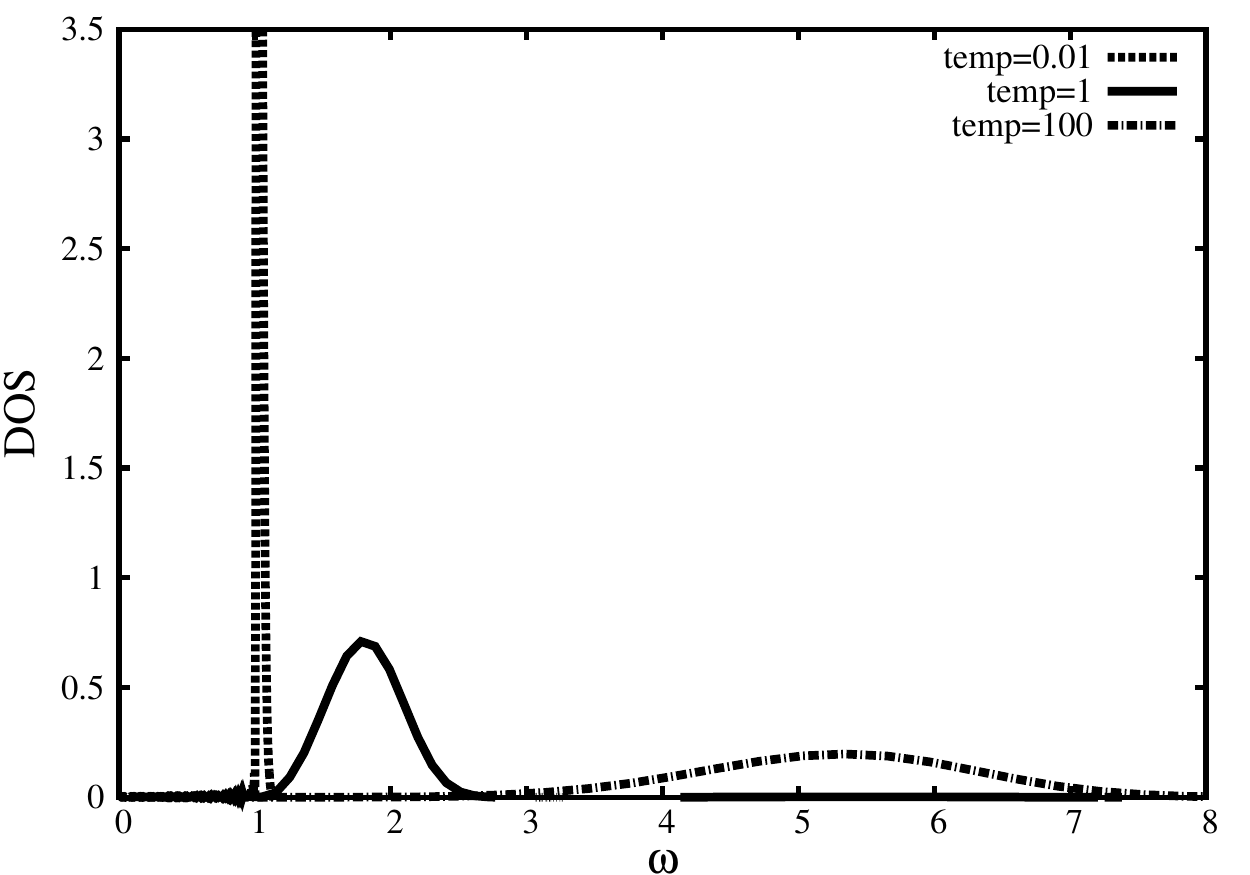}
\caption{Density of states at various temperatures for the $x^4$ model.} 
\label{DOS_x4}	
\end{figure}

As one might expect from the data-collapse, the DOS for the $x^4$
model is also simple, as shown in Fig~\ref{DOS_x4} for various
temperatures.
The DOS of this model is characterized by a single dominant peak that
shifts and broadens with temperature, as one would typically expect of
a vibrational mode in an anharmonic solid. In such a case, the
lifetime depends mostly on the shape of the DOS around the peak, and
two parameters (peak value of the DOS and the width) are sufficient to
describe it. At low temperatures, $\gamma_4 \rightarrow 1$, while at
high temperatures $\gamma_4 \rightarrow 2.2$ (for this model).
Recalling $\gamma_4$ as the (dimensionless) ratio $\mu_4/\mu_2^2$, it
is aptly designated as a ``shape parameter'' of the DOS.

\begin{figure}[H]
\centering
\includegraphics[scale=1.1]{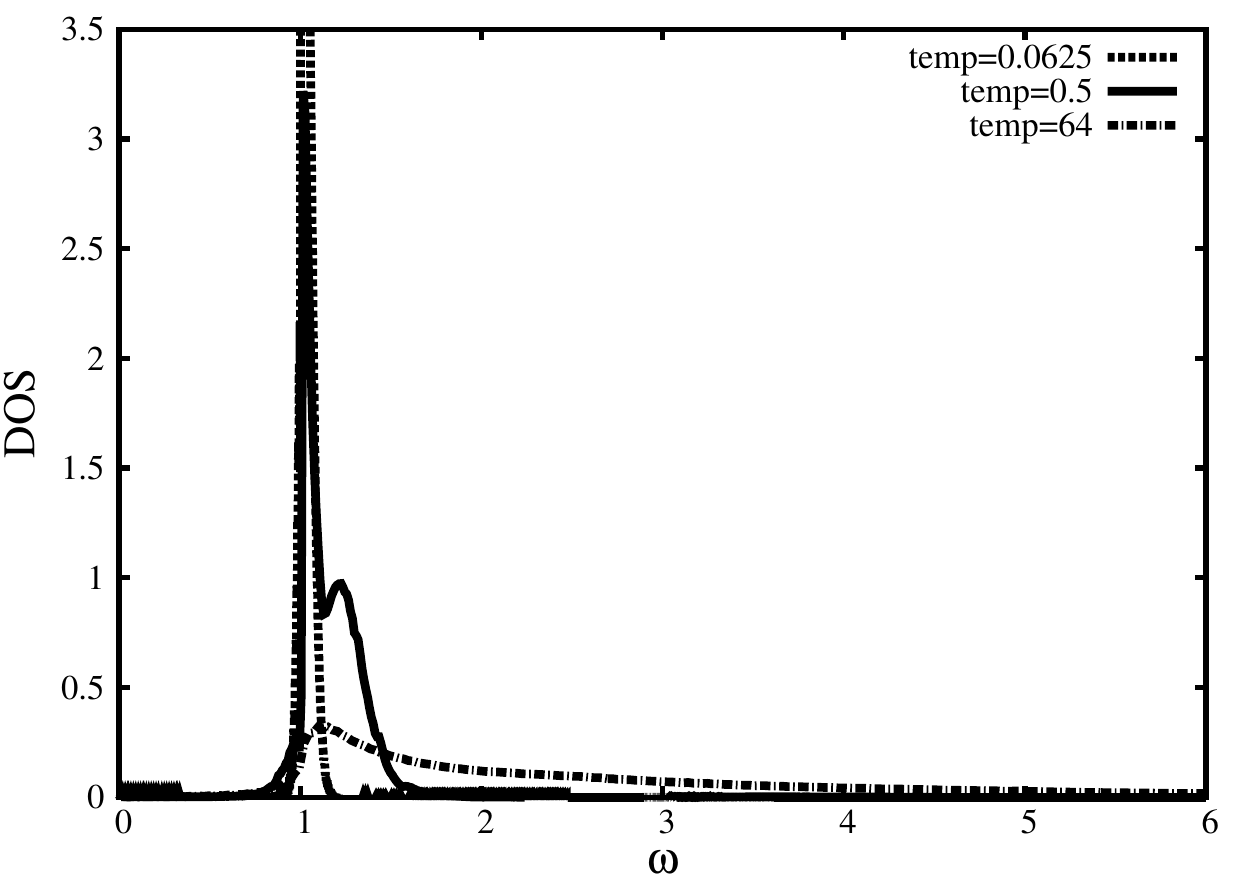}
\caption{Density of states of the $y$-mode at $M=1$ for the $x^2y^2$ model } 
\label{DOS_x2y2}	
\end{figure}

The DOS of the $x^2y^2$ model (Fig.~\ref{DOS_x2y2}) is only somewhat more
complex than that of the $x^4$ model. 

The simple evolution of the DOS with the temperature and other
parameters for these models explains why a simple, generic
relationship can exist between $\tau$ and the first two moments of the
DOS. To explore this point further, let us consider a generic,
single-mode DOS that is peaked at an oscillator frequency $\omega_0$
and broadened to a width $\Omega$. Both the oscillator frequency and
width will depend on temperature. At low temperatures, $\Omega <<
\omega_0$, and from Eq.~\ref{eq:n2} we have
\[ \tau \approx \Omega^{-1} \] The leading behavior of the lowest two
moments is
\[ \mu_2 \approx \omega_0^2 ( 1 + a \Omega^2/\omega_0^2 ) \]
\[ \mu_4 \approx \omega_0^4 ( 1 + b \Omega^2/\omega_0^2 ) \]
where $a$ and $b$ depend on the details of the DOS. Then
\[\tau/\tau_2 \approx \omega_0/\Omega \]
and 
\[\gamma_4 -1 \approx \Omega^2/\omega_0^2 \]
Eliminating $\Omega$ and $\omega_0$ among the two relations gives
\[ \tau/\tau_2 \approx (\gamma_4 - 1)^{-1/2} \] 
just as we found in Eq.~\ref{eqfit}. So long as the DOS has this
simple, generic behavior, the same relationship obtained here should
hold.

At high temperatures, if the DOS can be assumed to be a mostly
featureless and broad distribution with width $\Omega$ and height
$\Omega^{-1}$, then 
$ \tau \approx \Omega^{-1} $
and 
$ \mu_2 \approx \Omega^2 $
so 
$ \tau_2 \approx \Omega^{-1} $.
While the shape parameter saturates at some value ($ \gamma_4 \approx
c$), in which case the variation in $\tau$ is tracked by that of
$\tau_2$, so that
\[ \tau \approx \tau_2 \]
which is the behavior reported by DD.

\begin{figure}[H]
\centering
\includegraphics[scale=1.1]{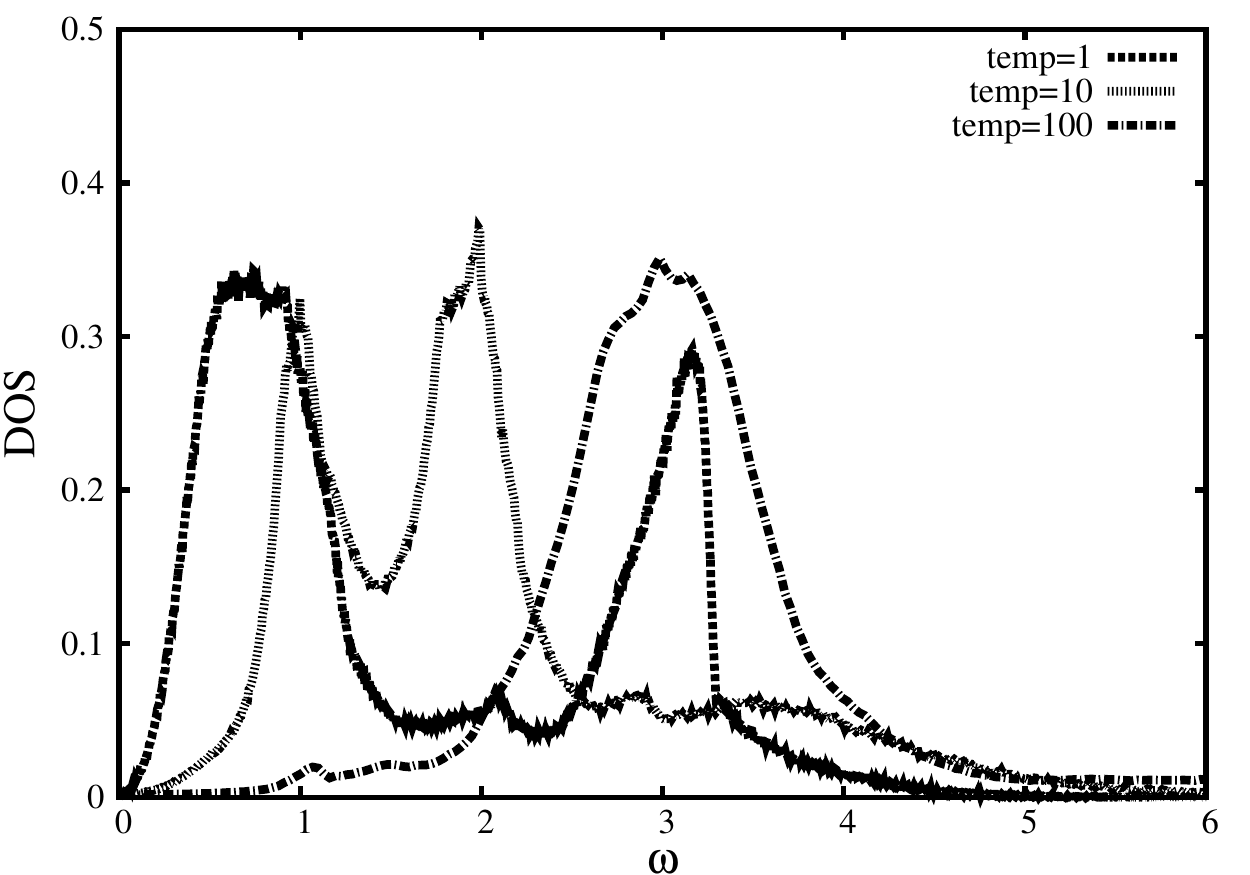}
\caption{Density of states of the $x$-mode at $\lambda=0.2$ and
  various temperatures for the cubic model.}
\label{DOS_cubic}	
\end{figure}

The DOS of cubic model (Fig.~\ref{DOS_cubic}) is much more complicated
than that of $x^4$ and $x^2y^2$ model, which explains why the simple
2-parameter scatterplot (Fig.~\ref{MACFcubic}) does not capture the
behavior.

Finally we note that $\gamma_4$, in addition to being a simple measure
of the shape of the DOS, is also a direct measure of the degree of
anharmonicity of the mode as averaged over the ensemble. Specifically,
$\gamma_4$ for a given mode can be re-written as
\begin{equation}
\gamma_4 = \frac{ \langle x^2 \rangle \langle f^2 \rangle }{ \langle x
f \rangle^2 }
\end{equation}
where $f$ is the force associated with a displacement $x$. A harmonic
system is, of course, defined where the force obeys $f+k x=0$. In the
anharmonic ensemble, we could define an effective $k$ by that which
minimizes the deviation from linear. That is, define the effective $k$
by minimizing $\alpha=\langle(f+kx)^2\rangle$. The minimum value of
$\alpha$ then measures the degree of anharmonicity of the system as
effective for the ensemble. For a harmonic system,
$\alpha_{\mathrm{min}}=0$. In general, $k_{\mathrm{eff}} = -
  \langle x f \rangle/\langle x^2 \rangle $ and
\begin{equation}
\alpha_{\mathrm{min}} = \frac{ \langle
  x f \rangle^2}{\langle x^2 \rangle} ( \gamma_4 -1 )
\end{equation}
showing how the deviation $\gamma_4-1$ is directly related to the
effective anharmonicity of the ensemble.

\section{Conclusions}

We have investigated using low-dimensional models the proposal that
the mode lifetime in equilibrium might be approximated from the two
lowest moments of the Liouvillian. For the generic case of a DOS
dominated by a single peak broadened and shifted, as is the case here
for the $x^4$ and $x^2 y^2$ models, we see that the fourth moment
approximation works well. In the case of the cubic model, the fourth
moment approximation is insufficient, which can be understood in terms
of the more complex structure of the DOS. The multiple minima of the
cubic model creates a more complex dynamics that cannot be captured
with just two parameters.

\section*{Acknowledgement}

Research supported by the U. S. Department of Energy, Office of Basic 
Energy Science, Division of Materials Sciences and Engineering under Award ER 46871.

\end{document}